\documentclass[%
 reprint,
superscriptaddress,
 amsmath,amssymb,
 aps,
 prx,
]{revtex4-2}

\usepackage{xcolor}
\usepackage{graphicx}
\usepackage{physics}
\usepackage{verbatim}
\usepackage[hidelinks]{hyperref}
\usepackage[mathlines]{lineno}
\usepackage{siunitx}
\usepackage{tabularx}

\usepackage[caption=false]{subfig}

\begin{document}

\title{Testing Single Photon Entanglement using Self-Referential Measurements}

\author{Daniel Kun}\email[Corresponding author: ]{daniel.kun@univie.ac.at}
\affiliation{University of Vienna, Faculty of Physics, Vienna Center for Quantum Science and Technology (VCQ)  Boltzmanngasse 5, 1090 Vienna, Austria}
\author{Teodor Strömberg}
\affiliation{Institute of Science and Technology Austria, Klosterneuburg, Austria}
\author{Borivoje Daki\'{c}}
\affiliation{University of Vienna, Faculty of Physics, Vienna Center for Quantum Science and Technology (VCQ)  Boltzmanngasse 5, 1090 Vienna, Austria}
\author{Philip Walther}\affiliation{University of Vienna, Faculty of Physics, Vienna Center for Quantum Science and Technology (VCQ)  Boltzmanngasse 5, 1090 Vienna, Austria}
\affiliation{Christian Doppler Laboratory for Photonic Quantum Computer, Faculty of Physics,  University of Vienna, 1090 Vienna, Austria}
\affiliation{Institute for Quantum Optics and Quantum Information (IQOQI) Vienna, Austrian Academy of Sciences, Boltzmanngasse 3, Vienna, Austria}
\author{Lee A. Rozema}\email[Corresponding author: ]{lee.rozema@univie.ac.at}
\affiliation{University of Vienna, Faculty of Physics, Vienna Center for Quantum Science and Technology (VCQ)  Boltzmanngasse 5, 1090 Vienna, Austria}

\date{November 26, 2025}

\begin{abstract}
Entanglement does not always require one particle per party. It was predicted some thirty years ago that a single photon traversing a beam splitter could violate a Bell inequality. Although initially debated, single-photon nonlocality was eventually demonstrated via homodyne measurements. Here, we present an alternate realisation that avoids the complexity of homodyne measurements and potential loopholes in their implementation. We violate a Bell inequality by performing joint measurements on two copies of the same single-photon entangled state, where one photon acts as a phase reference for the other, making it self-referential. We observe CHSH parameters of $2.71\pm 0.09$ and $2.23\pm 0.07$, depending on the joint measurements implemented. This offers a new perspective on single-photon nonlocality and a more accessible experimental route, potentially applicable to general mode-entangled states in diverse platforms.
\end{abstract}

\maketitle

\section{Introduction}
Quantum entanglement has been a topic of foundational interest for almost a century \cite{einstein1935can,schrodinger1935gegenwartige} and still leads to the discovery of surprising fundamental effects \cite{wang2025violation, Ciliberto2024Violation}. 
Recently, with the emergence of quantum information, entanglement has also become technologically relevant \cite{horodecki2009quantum}, underpinning a multitude of novel applications in quantum technologies\cite{ekert1991quantum,jozsa1997entanglement,bartolucci2023fusion,demkowicz2014using,lo2012measurement}.
In the simplest case of bipartite entanglement, two parties, Alice and Bob, each possess half of an entangled system.
By each measuring their part of the system, they reveal nonlocal correlations, evidenced by their ability to violate a Bell inequality \cite{bell1964einstein,clauser1969proposed}.
In this case, it is perhaps most natural to assume that the system consists of a pair of entangled particles, with each party measuring one of the particles.
Indeed, this is the typical setting for demonstrations and applications of entanglement.
For example, in experimental studies of bipartite entanglement, Alice and Bob are often each given one photon from an entangled \textit{photon pair} and measure one of its internal degrees of freedom, such as the photon's polarization.
However, bipartite quantum systems do not require two particles.
In particular, in the early 1990s, an idea for demonstrating a Bell violation using
\textit{single-particle entanglement} was first proposed \cite{oliver1989predictions,tan1991nonlocality,yurke1992bell}. More recently, single-particle entanglement has reemerged in the context of applications in quantum information \cite{alonso2019trace,del2018two,massa2019experimental,massa2022experimental,BozzioQuantum2020,steffinlongo2024long,kun2025direct, wang2025memory} due to its relative experimental ease. 

\begin{figure}[ht]
    \centering
    \includegraphics[width=\linewidth]{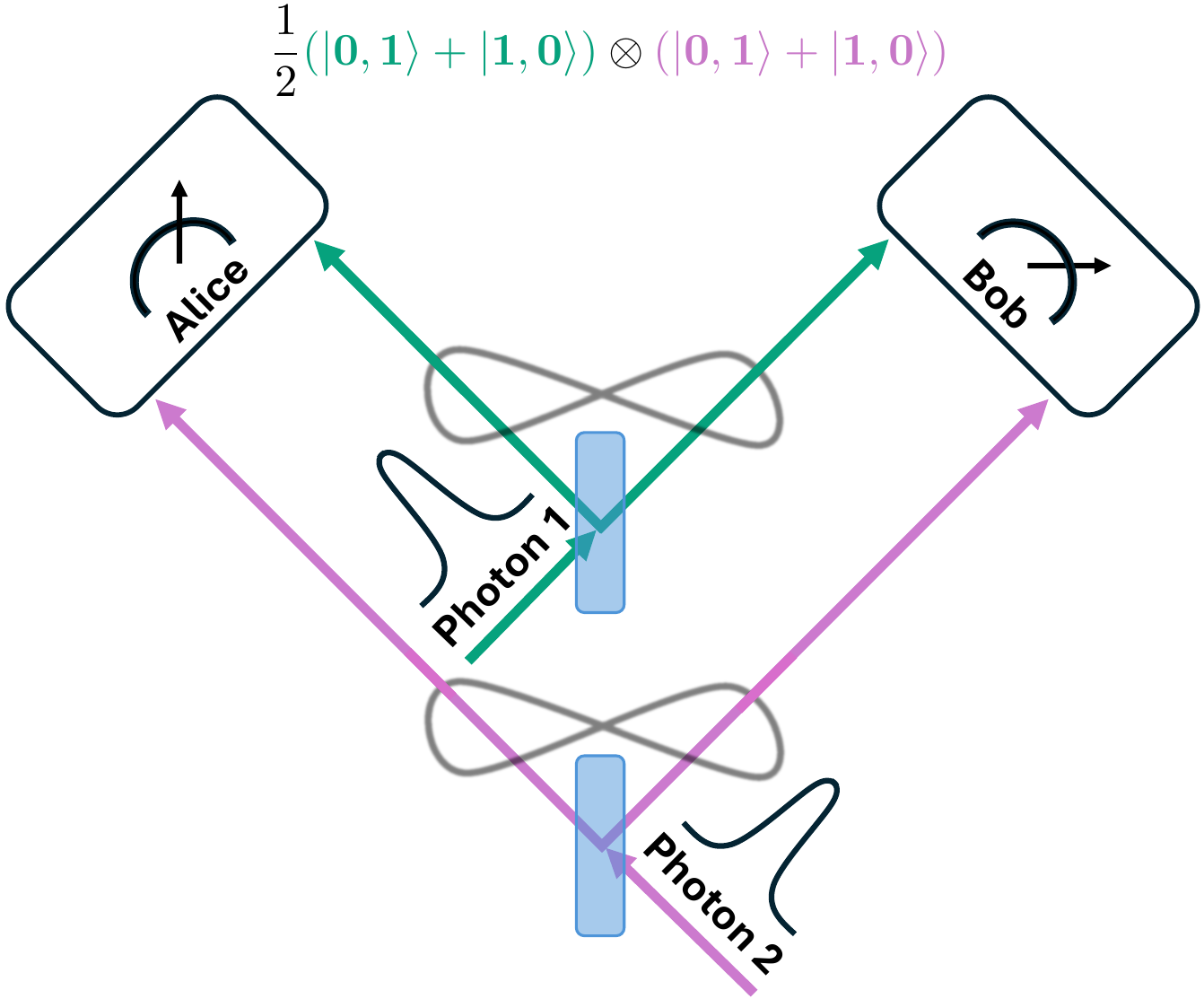}
    \caption{\textbf{Self-referential measurements}. 
    Our scheme requires two identical copies of a single-photon entangled state.
    Each copy is produced by sending a single photon to a beam splitter.
    The two modes of each single-photon entangled state are then shared between Alice and Bob, which they can use to violate a CHSH inequality by performing joint measurements on them.
    Prior to these joint measurements, there is no entanglement between photon 1 and photon 2.
    In this scheme, photon 2 can be seen to act as a phase reference for photon 1, and vice versa, making this a self-referential measurement.
    }
    \label{fig:concept}
\end{figure}

In single-photon entanglement, one photon is sent to a 50/50 beam splitter (BS) at a central location after which one output mode of the BS is sent to Alice and the other to Bob.
The photon is thus placed in a superposition of {traveling to} Alice or Bob; \textit{i.e.} it is in the entangled state $\ket{\psi} = \frac{1}{\sqrt{2}}(\ket{0, 1}_{A,B} + \ket{1, 0}_{A,B})$
(see the green modes in Fig.~\ref{fig:concept}).
Despite initially sparking a debate \cite{hardy1994nonlocality, greenberger1995nonlocality,vanenk2005single}, the Bell-nonlocality of single particles was eventually established experimentally \cite{fuwa2015experimental,guerreiro2016demonstration}. 
As there is now only a single particle involved, one can view the ``entangled entities'' as the two modes that are entangled in their particle number. 
Such scenarios are often referred to as \textit{mode-entanglement} \cite{hillery2006entanglement}. 
The so-called N00N states are one prominent example of mode-entanglement \cite{boto2000quantum,dowling2008quantum}. 
While it is well known that N00N states can enhance the resolution and sensitivity of measurements \cite{mitchell2004super,rozema2014scalable,rozema2014optimizing} it has also been proposed that they can violate Bell inequalities \cite{Wildfeuer2007Strong}.
Correlations in the number basis of mode-entangled states (such as the single-photon example described above) are clear: if Alice finds the photon in her lab, Bob will measure vacuum, and vice versa.
However, violating a Bell inequality requires Alice and Bob to measure correlations in different bases; \textit{i.e.} in a particle-number superposition basis.

In photonic settings, the required measurements are typically achieved using homodyne detection \cite{yukawa2013generating,Wildfeuer2007Strong}.
These measurements are known to be technically complex and are especially difficult to apply to mode-entangled states consisting of massive particles \cite{ferris2008detection}. 
In a homodyne measurement, Alice and Bob must share a common phase reference, called a local oscillator (LO), which they locally interfere with their respective single-photon modes.
Experimentally, the shared LO is created by splitting a laser beam (which is nominally a coherent state) on a BS at a central node and sending the two split beams to Alice and Bob. 
This approach to creating the shared LO opens up an objection stemming from a long-running debate about the nature of coherence in quantum optics\cite{molmer1997optical,rudolph2001requirement,bartlett2006dialogue}, in which the LO itself can be argued to carry nonlocal correlations after being split on the BS. 
Therefore, this raises questions about both the practicality and the interpretation of experiments demonstrating single-particle nonlocality using homodyne measurements.

In this article, we present an alternative approach to the violation of a Bell inequality using single-photon entangled states. As in some past proposals \cite{yurke1992bell}, our scheme does not rely on a shared phase reference or homodyne measurements. 
Instead, two copies of a single-photon entangled state are shared, and Alice and Bob then perform joint measurements on their respective modes (Fig.~\ref{fig:concept}). 
In these measurements, each photon can be thought of as a phase reference for the other photon, making the measurements \textit{self-referential}. 
Below, we derive a Bell inequality for our experimental scenario and show that Alice and Bob can use relatively simple joint measurements to violate it. 
Since the only resource in our experiment is two copies of the same single-photon entangled state, the violation of the inequality unambiguously demonstrates the Bell nonlocality of single photons at a beam splitter.

\section{Protocol}
Our experiment begins with two independent but identical photons $\gamma_1$ and $\gamma_2$.
Each photon is sent to an individual and separate BS at a central node. This generates two copies of the single-photon entangled state:
\begin{align}\label{eq:input-state}
    \ket{\psi_{\mathrm{in}}} = &\frac{1}{2}(\ket{0, 1}_{A_1B_1} + \ket{1, 0}_{A_1B_1}) \nonumber \\
    &\otimes (\ket{0, 1}_{A_2B_2} + \ket{1, 0}_{A_2B_2}). 
\end{align}
One mode of $\gamma_1$ and one mode of $\gamma_2$ are sent to Alice (modes $A_1$ and $A_2$, respectively), while the other two modes are sent to Bob ($B_1$ and $B_2$).
It is important to note that, at this point, there is no entanglement between photons $\gamma_1$ and $\gamma_2$.

A phase shifter is placed in one of the two modes entering both Alice's and Bob's labs. 
This phase shifter is used to implement a basis choice $\phi_x$ for Alice and $\phi_y$ for Bob; each party can set one of two different phases, making $x$ and $y$ binary basis choice variables. 
Alice (Bob) then performs a joint measurement on modes $A_1$ and $A_2$ ($B_1$ and $B_2$), locally interfering them on a BS and placing {number-resolving} detectors $D_{A1}, D_{A2}$ ($D_{B1}, D_{B2}$) in the two output ports.
We assume that the quantum state across the modes $A_1$ and $B_1$ is identical to the state in $A_2$ and $B_2$.

Once the bases have been set and the two modes have interfered in Alice's and Bob's labs, there are ten different possible detection patterns between Alice and Bob, which are summarized in the left column of Tab.~\ref{tab:output-patterns}. 
Both photons can be recorded at a single detector, resulting in a \textit{double click} (4 patterns); one photon can be detected in each detector within the same lab, resulting in an \textit{in-lab coincidence} (2 patterns); or one photon can be registered in each lab, yielding a \textit{cross-lab coincidence} (4 patterns). Note that when one party detects a double click event in their lab, the other party must detect vacuum.

\begin{table}[ht]
\centering
\small
\setlength{\tabcolsep}{3pt}
\renewcommand{\arraystretch}{1.5}
\begin{tabularx}{\columnwidth}{l X c}
\hline
Click pattern & Probability $P_{N_{A1}N_{A2},N_{B1}N_{B2}}$ & Outcome \\ 
\hline
Alice double click 
  & $P_{20,00} = P_{02,00} = \tfrac18$ 
  & $(-,+)$ \\

Bob double click 
  & $P_{00,20} = P_{00,02} = \tfrac18$ 
  & $(+,-)$ \\

In-lab (HOM dip)
  & $P_{11,00} = P_{00,11} = 0$ 
  & $(+,+)$ \\

\hline

Cross-lab $A_1,B_1$ 
  & $P_{10,10} = \frac{1}{4} \sin^2\left(\frac{\phi_x+\phi_y}{2}\right)$
  & $(-,-)$ \\

Cross-lab $A_2,B_2$ 
   & $P_{10,10} = \frac{1}{4} \sin^2\left(\frac{\phi_x+\phi_y}{2}\right)$
  & $(+,+)$ \\

Cross-lab $A_1,B_2$ 
   & $P_{10,10} = \frac{1}{4} \cos^2\left(\frac{\phi_x+\phi_y}{2}\right)$
  & $(-,+)$ \\

Cross-lab $A_2,B_1$ 
   & $P_{10,10} = \frac{1}{4} \cos^2\left(\frac{\phi_x+\phi_y}{2}\right)$
  & $(+,-)$ \\

\hline
\end{tabularx}
\caption{Detection patterns for Alice and Bob and their quantum--mechanical probabilities}
\label{tab:output-patterns}
\end{table}

\begin{figure}[ht]
\centering
 \includegraphics[width=\linewidth]{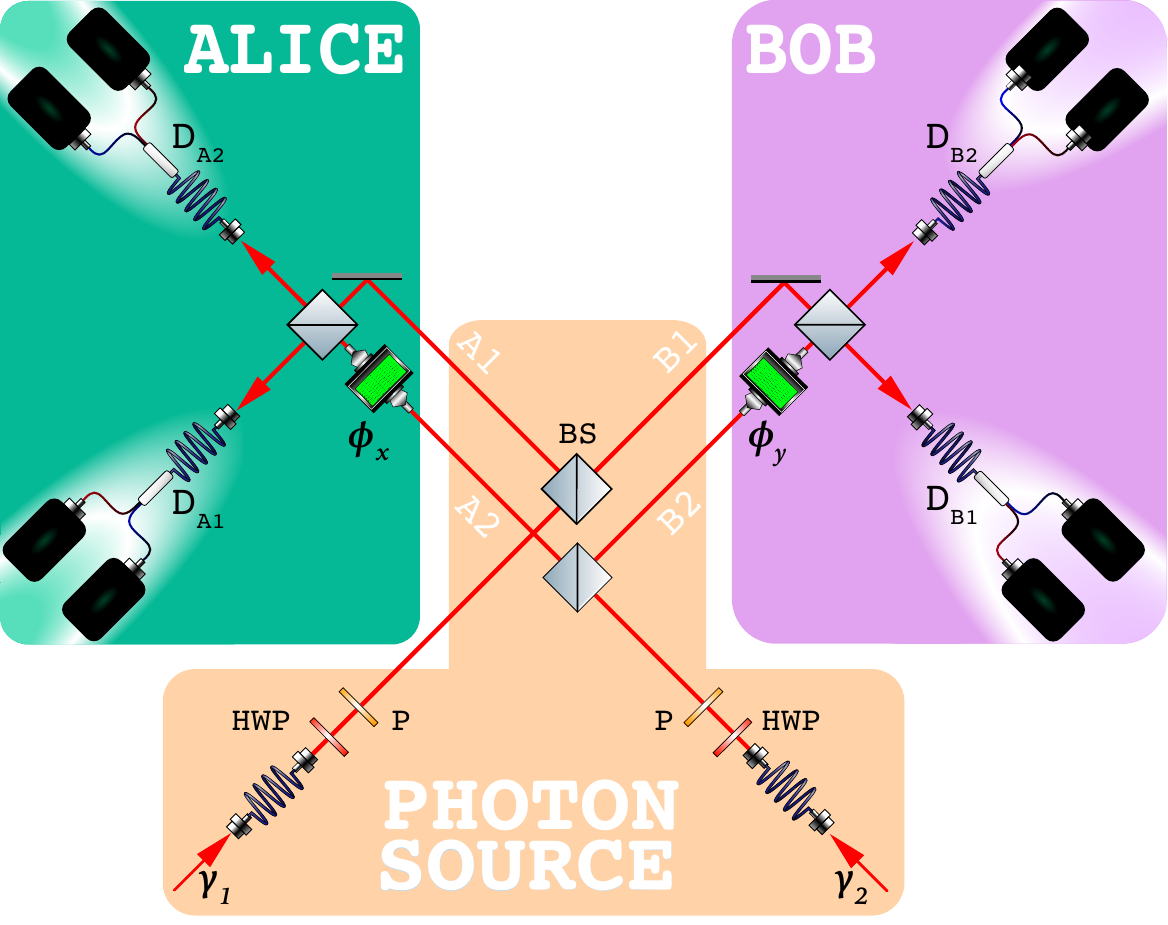}
\caption{
\textbf{Experimental Setup.} A source (orange field) generates spectrally identical photons $\gamma_1$ and $\gamma_2$, which are made indistinguishable in polarization by passing through a half-wave plate (HWP) and a linear polarizer (P) and are then individually placed in an entangled state using beam splitters (BSs). The two {entangled states} are then shared between Alice's lab (green field) and Bob's lab (magenta field) who each locally control a piezo-controlled phase shifter $(\phi_x, \phi_y)$, which implement their basis choices. Alice and Bob then perform joint measurements on their two photon modes, locally interfering them on a BS and using pseudo--number-resolving detectors ($D_{A_{1/2}}, D_{B_{1/2}}$) consisting of a fiber beam splitter and two detectors each, allowing them to violate a CHSH inequality.
}
\label{fig:optical-setup}
\end{figure}

In order to adhere to the well-known CHSH paradigm \cite{clauser1969proposed}, Alice and Bob each attribute $+1$ or $-1$ to their detection events. Their possible joint outcome values (\textit{outcome groups}) are shown in the right column of Tab.~\ref{tab:output-patterns}.
Since each party has two measurement choices (Alice applies $\phi_{x=0}$ or $\phi_{x=1}$ and Bob applies $\phi_{y=0}$ or $\phi_{y=1}$) and two possible outcomes ($+1$ or $-1$), the usual CHSH inequality holds for any local hidden variable theory:
\begin{equation}\label{eq:chsh-defn}
    |\mathcal{B}| = |\left< C_{00} \right>  + \langle C_{01} \rangle + \langle C_{10} \rangle - \langle C_{11} \rangle| \leq 2.
\end{equation}
In Eq.~(\ref{eq:chsh-defn}) we have defined the correlation parameters
\begin{equation}\label{eq:corr-param-def}
C_{xy} = P_{++} + P_{--} - P_{+-} - P_{-+}
\end{equation}
between Alice and Bob for each of their basis choices $(x, y)$, with the probabilities $P_{ab}$ defined as below.

We now compute the maximal value $|\mathcal{B}|$ can reach in our experiment, assuming quantum mechanics. 
The probabilities for each of the ten detection patterns are given in the middle column of Tab.~\ref{tab:output-patterns}, where $N_{A_1}$ stands for the number of photons registered at Alice's detector $D_{A_1}$, etc.
Notice that the probability for an in-lab coincidence (row 3) is nominally $0$.
This requires the assumption that the two photons are perfectly indistinguishable, such that if two photons enter the same lab, they will undergo Hong-Ou-Mandel interference and bunch, resulting in a double click event.
The resulting probabilities for each outcome group, according to Tab.~\ref{tab:output-patterns}, are thus
\begin{align}\label{eq:nps-group-probs}
P_{++} &= P_{11,00} + P_{00,11} + P_{01,01} = \frac{1}{4} \sin^2 \phi_{xy} \nonumber \\
P_{+-} &= P_{00,20} + P_{00,02} + P_{01,10} = \frac{1}{4} (1 + \cos^2 \phi_{xy}) \\
P_{-+} &= P_{20,00} + P_{02,00} + P_{10,01} = \frac{1}{4} (1 + \cos^2 \phi_{xy}) \nonumber \\
P_{--} &= P_{10,10} = \frac{1}{4} \sin^2 \phi_{xy} \nonumber,
\end{align}
where we have defined the average phase $\phi_{xy}= \frac{\phi_x+\phi_y}{2}$
determined by Alice's and Bob's basis choices $(x, y)$.\\
\\
In order to detect all possible click patterns from Tab.~\ref{tab:output-patterns}, we must use number-resolving detectors, since standard click detectors cannot register double clicks and, thus, implement a post-selection on events where the photons arrive at different detectors.
In the case of standard click detectors, half of the occurring events (top two rows) are discarded, and the remaining events are renormalized, resulting in a $\frac{1}{2}$ pre-factor instead of $\frac{1}{4}$ such that the new joint probabilities read:
\begin{align}
P_{++} &= P_{11,00} + P_{00,11} + P_{01,01} = \frac{1}{2} \sin^2 \phi_{xy}  \nonumber \\
P_{+-} &= P_{01,10} = \frac{1}{2} \cos^2 \phi_{xy} \\ 
P_{-+} &= P_{10,01} = \frac{1}{2} \cos^2 \phi_{xy} \nonumber \\
P_{--} &= P_{10,10} = \frac{1}{2} \sin^2 \phi_{xy}. \nonumber
\end{align}
Thus, without number-resolution, the correlation parameter from Eq.~(\ref{eq:corr-param-def}) can be written as  $\tilde{C}_{xy} = - \cos 2\phi_{xy}$ and the CHSH parameter from Eq.~(\ref{eq:chsh-defn}) takes the form
\begin{equation}\label{eq:bell-ps}
|\mathcal{\tilde{B}}| = \left| \cos 2\phi_{00} + \cos 2\phi_{01} + \cos 2\phi_{10} - \cos 2\phi_{11} \right|,
\end{equation}
which can be simplified to
\begin{equation}\label{eq:chsh-ps}
    |\mathcal{\tilde{B}}| = |3 \cos 2\theta - \cos 6\theta|
\end{equation}
by choosing the bases to be $\phi_{00} = \theta, ~~ \phi_{01} = \phi_{10} = -\theta, ~~ \phi_{11} = -3\theta$.
The maximal violation of Eq.~(\ref{eq:chsh-defn}) occurs for $\theta = \pi/8$, yielding $|\mathcal{\tilde{B}}| = 2\sqrt{2}$, which is achieved with individual phase settings $\phi_{x=0} = \pi/8, ~ \phi_{x=1} = -3\pi/8, ~ \phi_{y=0} = \pi/8$ and $\phi_{y=1} = -3\pi/8$, for example.
Thus, by neglecting the double click events, we can violate the CHSH inequality up to the Tirelson bound.\\
\\
If, instead, we use number-resolving detectors, we gain access to all detection patterns in Tab.~\ref{tab:output-patterns} and the correlation parameter becomes $C_{xy} = -\cos^2 \phi_{xy}$,
which in turn yields the non--post-selected CHSH parameter 
\begin{equation}\label{eq:chsh-nps}
    |\mathcal{B}| = \left|3 \cos^2 \theta - \cos^2 3\theta \right|
\end{equation}
for the same basis choices as above. This new expression for the CHSH parameter yields a maximal value of $|\mathcal{B}| = 1 + \sqrt{2}$, which is greater than 2 but lower than $|\mathcal{\tilde{B}}|$ obtained with post-selection.
The reduction in the maximum achievable violation arises because the double click events are independent of the phases set by Alice and Bob (see Tab.~\ref{tab:output-patterns}), and these events thus only contribute classical correlations.
Half of all detection events are double clicks, making it somewhat surprising that it is possible to violate the CHSH inequality while still including these terms.
In both cases, this measurement scheme reveals the presence of nonlocal correlations between Alice and Bob by simply using two copies of the same single-photon entangled state.

\begin{figure*}[ht]
    \centering
    \includegraphics[width=.9\linewidth]{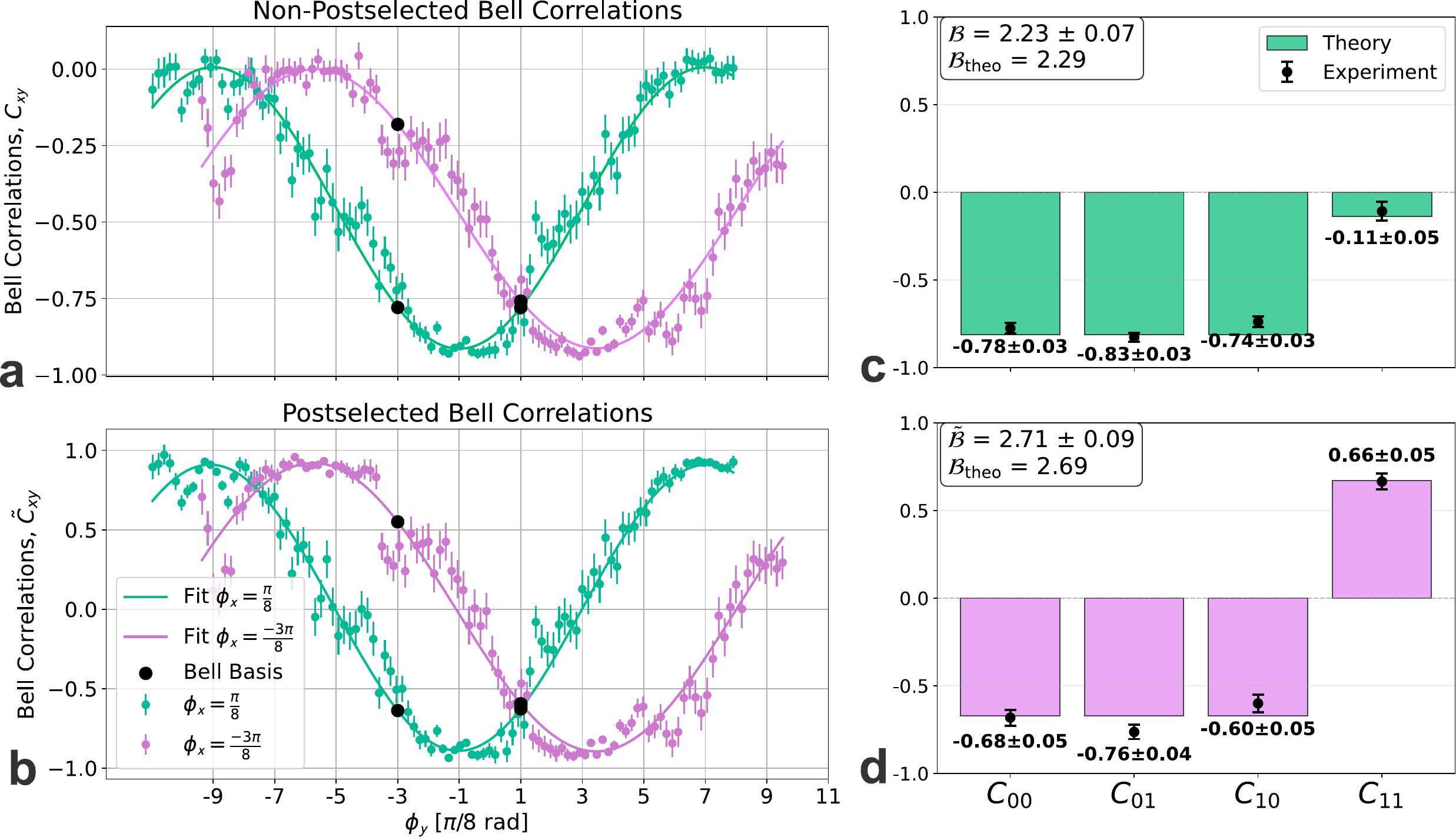}
    \caption{\textbf{Bell Correlations}. We measure the visibility of the Bell correlations, fixing Alice's phase at $\pi/8$ (green) and $-3\pi/8$ (magenta) -- corresponding to the two basis settings to maximize the CHSH parameter -- and scan Bob's phase. The Bell correlations oscillate with an amplitude of \textbf{a.} $0.902 \pm 0.011$ when not post-selecting and \textbf{b.} $0.914 \pm 0.012$ with post-selection, slightly lower than the measured HOM visibility. The black points indicate the location of the phase set points for Alice's and Bob's Bell basis choices $\phi_x, \phi_y \in \{\pi/8, -3\pi/8\}$ determined during calibration measurements similar to these data. \textbf{c.} The final Bell correlations and Bell parameters are determined in subsequent measurements setting the previously calibrated voltage values of the black points and measuring only these four Bell bases as opposed to the full scan. The estimated Bell correlations in each basis setting are compared to the theoretical predictions from Eq.~(\ref{eq:viz-corr-params})
    without post-selection and \textbf{d.} with post-selection.}
    \label{fig:bell-visibility}
\end{figure*}

\section{Experimental Details and Results}
We experimentally implement our protocol using single photons, as sketched in Fig.~\ref{fig:optical-setup}.
Pairs of indistinguishable single photons ($\gamma_1, \gamma_2$) are generated using a type-II beta barium borate (BBO) spontaneous parametric down-conversion (SPDC) source. In order to maximize the spectral indistinguishability, the photons are passed through \SI{3}{\nano\meter} bandpass filters before being sent through a half wave-plate (HWP) and a linear polarizer (LP) to ensure indistinguishability in the polarization degree of freedom. Next, each photon is individually prepared in a mode-entangled state by sending it through a 50/50 beam splitter, which creates the input state $\ket{\psi_{\mathrm{in}}}$. 

\begin{figure}[ht]
    \centering
    \includegraphics[width=\linewidth]{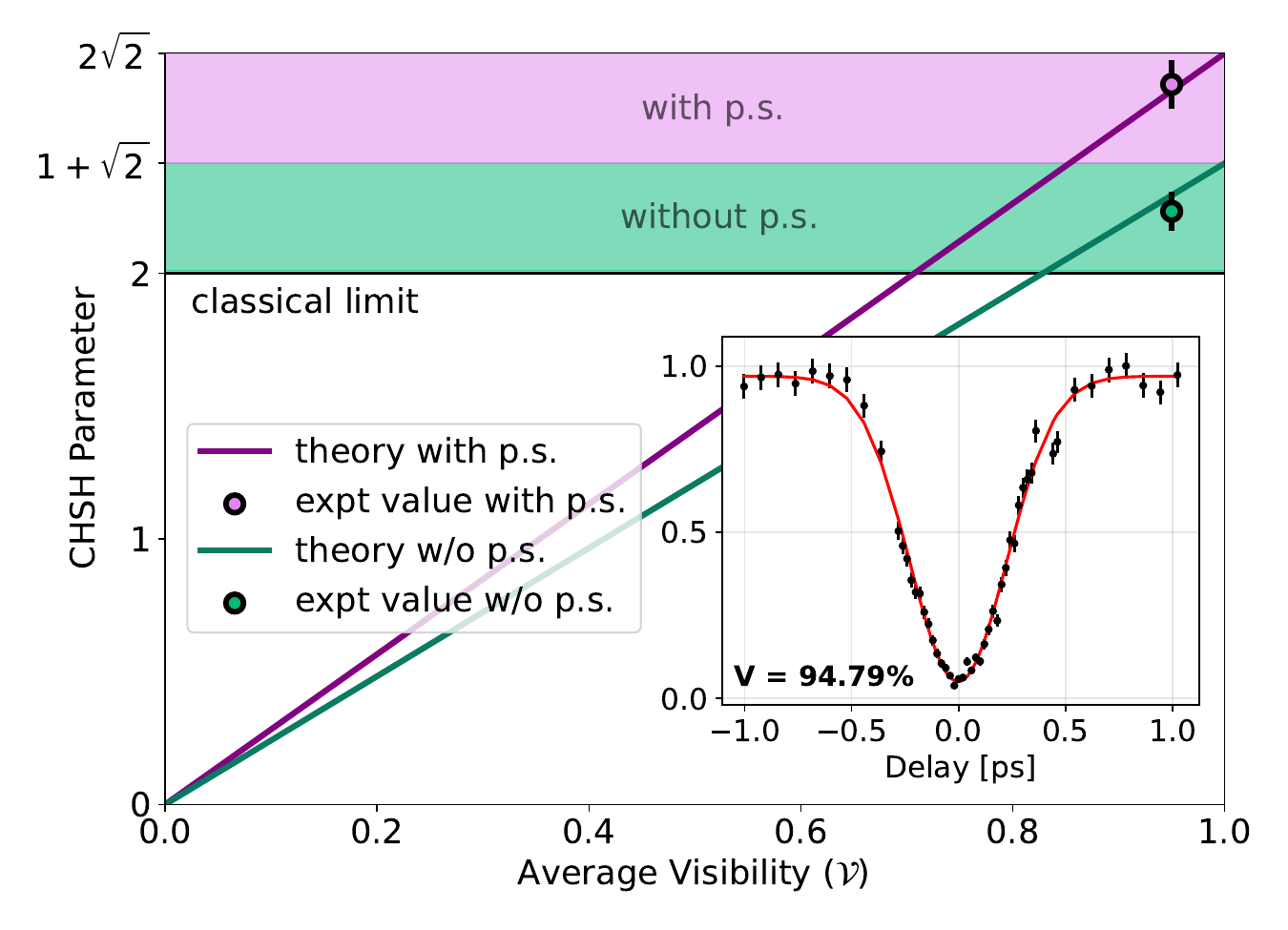}
    \caption{
    Alice and Bob compute the two-copy CHSH parameter by comparing their detections for their four basis choices. The maximally achievable parameter value depends linearly on the indistinguishability of the two photons. For perfect distinguishability, the Hong-Ou-Mandel (HOM) visibility is 1 and the maximal parameters are $1+\sqrt{2}$ without post-selection and $2\sqrt{2}$ with post-selection. We achieved an average visibility of \SI{95}{\percent} and a respective violation of $\mathcal{B} = 2.23 \pm 0.07$ and  $\mathcal{\tilde{B}} = 2.71 \pm 0.09$, which is in good agreement with the theoretically achievable values of 2.293 and 2.687.
    }
    \label{fig:bell-violation-hom}
\end{figure}

Since the success of the protocol fundamentally relies on the indistinguishability of the photon pairs, we measure the Hong-Ou-Mandel (HOM) visibility \cite{branczyk_hong-ou-mandel_2017} in both Alice and Bob's labs, which we define as
\begin{equation}
    \mathcal{V}_A = 1 - \frac{P_{11,00}}{P_{20,00} + P_{02,00}}, ~\mathcal{V}_B = 1 - \frac{P_{00,11}}{P_{00,20} + P_{00,02}},
\end{equation}
respectively. 
We do so by scanning the temporal delay between the two photons. The in-lab coincidences (row 3 in Tab.~\ref{tab:output-patterns}) vanish for perfectly indistinguishable particles and thus serve as an experimental proxy for indistinguishability. We find an average visibility of $\mathcal{V} = \SI{95.2(6)}{\percent} $  based on a Monte Carlo simulation of the Gaussian fit of the HOM dip scan (see Fig.~\ref{fig:bell-violation-hom} inset).

To resolve two-photon bunching events, we place a fiber-based beam splitter together with a pair of single-photon detectors in each output port of Alice's and Bob's bulk beam splitters, as depicted in Fig.~\ref{fig:optical-setup}. This yields an effective two-photon detection efficiency of \SI{50}{\percent}, allowing us to detect the double-click events, corresponding to row 3 in Tab.~\ref{tab:output-patterns}.

Before each experimental run, we measure the relative efficiency of each coincidence channel by setting the temporal delay between the two photons to be far greater than their coherence length. This renders the photons perfectly distinguishable, such that all coincidence patterns become equally likely. We can then straightforwardly assess the relative efficiency of each channel by directly comparing all coincidence rates. Together with the Poissonian uncertainty of the SPDC source and the phase noise, the detector efficiencies are the dominant contributors to the errors we quote below.

While the HOM visibility is a direct measurement of the indistinguishability of the two photons, the Bell visibility is a measure of how strongly we can hope to violate a Bell inequality with them. Modeling our experiment with partial distinguishability ($\mathcal{V} < 1$) similarly to the approach in \cite{kun2025direct} we can easily see that, for imperfect visibility, the correlation parameters above become
\begin{equation}\label{eq:viz-corr-params}
    \tilde{C}_{xy} = -\mathcal{V}\cos 2\phi_{xy} \quad \text{and} \quad 
    C_{xy} = -\mathcal{V}\cos^2 \phi_{xy} 
\end{equation}
respectively, for the post-selected and non--post-selected cases. In the absence of any further experimental imperfections, the Bell visibility should equal the HOM visibility, as indicated by Eq. (\ref{eq:viz-corr-params}). We measure the Bell correlation visibilities by first fixing Alice's phase basis to one ($\phi_x = \pi/8$) and then the other ($\phi_x = -3\pi/8$) Bell phase basis, sweeping Bob's phase across a full fringe for each and acquiring coincidence counts for \SI{2.5}{s} per data point. The sweep measurement lasts approximately \SI{5}{min}, which is on the order of the phase stability. We then evaluate the Bell correlations with and without post-selection for each data point and fit the resulting data with the expressions from Eq. (\ref{eq:viz-corr-params}), extracting the amplitudes from the fit.
We find a Bell correlation visibility of \SI{91.4(1.8)}{\percent} (\SI{90.2(1.2)}{\percent}) {without (with)} post-selection, which is slightly below the HOM visibility. Due to the long acquisition time of the correlation sweep, we observe a slight chirp in the sinusoidal data, indicating a slight phase drift that affects the fitted amplitude. Furthermore, we assumed a phase noise of \SI{2}{\percent} of the length of a fringe, which leads to increased uncertainty in the computed Bell correlations at the steep sections of the curve, further reducing the fitted amplitude.

In order to measure the final CHSH parameter, we use the aforementioned phase sweeps as a calibration run, but with significantly shorter acquisition times (\SI{300}{\milli\second}) prior to the experimental run. 
We extract the four basis phase set points from these phase-sweep data for Alice and Bob (black dots in Fig.~\ref{fig:bell-visibility} are representative points). During each data run, we take 30 data points with \SI{1}{s} acquisition time for each basis choice to compute the CHSH parameter bringing each CHSH data run to a total time of approximately \SI{5}{min}, including \SI{2}{min} for phase calibration sweeps and \SI{30}{s} for efficiency calibration, comparable to the duration of phase stability. 
We find a CHSH parameter of $\mathcal{B} = 2.23 \pm 0.07$ without post-selection and $\mathcal{\tilde{B}} = 2.71 \pm 0.09$ when post-selecting on distributed cases only, where the overall uncertainty includes contributions from photon number uncertainty, phase noise, detector efficiencies, and fitting errors in the phase sweeps.
In Fig.~\ref{fig:bell-violation-hom}, the maximum theoretically achievable values for $\mathcal{B}$ are plotted against HOM visibility ($\mathcal{V}$), and the areas of violation of the classical bound without post-selection (green) and with post-selection (magenta) are indicated. Our experimentally achieved values match the theoretical simulation and demonstrate how a Bell inequality can be violated using two copies of a single-photon entangled state.

\section{Discussion}
In this work, we have presented a violation of a Bell inequality using single-photon entangled states. 
We emphasize that while each run of our experiment utilizes two copies of a single-photon state, these states do not share any initial entanglement. 
Rather, we use these two copies as phase references for one another, making our measurements self-referential.
There are many other interesting mode-entangled states with higher-photon number, including symmetric states \cite{goldberg2022quantumness} such as polarization-squeezed states \cite{shalm2009squeezing}, 
N00N states \cite{Wildfeuer2007Strong},
Platonic states \cite{ferretti2024generating}, as well as other general 2-mode states \cite{hillery2006entanglement}, for which similar techniques could apply.
In this context, our experiment can also be viewed as a violation of a Bell inequality using two copies of a single-photon N00N state.
Our self-referential technique could be particularly advantageous for mode-entangled states made up of massive particles, for which the use of homodyne measurements is not straightforward
\cite{ferris2008detection}, and it can remove the additional technical overhead of implementing homodyne measurements in photonic systems.

To the best of our knowledge, our protocol does not introduce any new loopholes when compared to homodyne schemes.
In fact, it allows one to address a critique of previous demonstrations of single-photon nonlocality based on homodyne measurements \cite{fuwa2015experimental,guerreiro2016demonstration}, which require the two parties to have distributed LOs with a shared phase reference between their labs. 
In textbook quantum optics, this phase reference consists, simply, of two in-phase coherent states, which is a separable quantum state that does not carry nonlocal correlations.
However, experimentally, this scenario is achieved by splitting a single laser beam on a beam splitter.
Ref.~\cite{rudolph2001requirement} pointed out that this does not result in a separable quantum state and that, when such a phase reference is used, the correlations can be equally well understood as originating in the shared reference itself \cite{molmer1997optical,rudolph2001requirement,bartlett2006dialogue}. 
Since there are only single-photon entangled states in our experiment, with no entanglement between these two photons before they enter Alice and Bob's labs, this can be considered a conclusive test of single-photon nonlocality.

\begin{acknowledgements}
This project has received funding from the European Union (ERC, GRAVITES, No 101071779), the European Union’s Horizon 2020 research and innovation programme under grant agreement No 899368 (EPIQUS), the European Union’s Horizon 2020 research and innovation programme under the Marie Skłodowska-Curie grant agreement No 956071 (AppQInfo) and the European Union (HORIZON Europe Research and Innovation Programme, EPIQUE, No 101135288). 
This research was funded in whole or in part by the Austrian Science Fund (FWF)[10.55776/COE1] (Quantum Science Austria), [10.55776/F71] (BeyondC) and [10.55776/FG5] (Research Group 5). 
This material is based upon work supported by the Air Force Office of Scientific Research under award number FA9550-21-1-0355 (Q-Trust) and FA8655-23-1-7063 (TIQI).

Views and opinions expressed are however those of the author(s) only and do not necessarily reflect those of the European Union or the European Research Council Executive Agency.
For open access purposes, the author has applied a CC BY public copyright license to any author accepted manuscript version arising from this submission.
The financial support by the Austrian Federal Ministry of Labour and Economy, the National Foundation for Research, Technology and Development and the Christian Doppler Research Association is gratefully acknowledged.
\end{acknowledgements}

\bibliography{nlbell_biblio}

\end{document}